\documentclass[aps,pra,preprint,groupedaddress,showpacs]{revtex4}

\usepackage{graphicx}
\usepackage{dcolumn}
\usepackage{bm}

\newcommand{\meanI}[1]{\langle #1 \rangle}
\newcommand{\var}[2]{\meanI{(\Delta \hat{#1}_{#2})^{2}}}
\newcommand{\hatn}[2]{\hat{#1}_{#2}^{\dag}\hat{#1}_{#2}}
\newcommand{\hatnc}[3]{\hat{#1}_{#2}^{\dag}\hat{#1}_{#3}}

\begin{document}

\title{Bichromatic Local Oscillator for Detection of Two-Mode Squeezed States of Light}

\author{Alberto M. Marino}
\email{marino@optics.rochester.edu}
\author{C. R. \surname{Stroud},~Jr.}
\author{Vincent Wong}
\altaffiliation[Current address: ]{DSO National Laboratories, 20
Science Park Drive, 118230, Singapore}
\author{Ryan S. Bennink}
\altaffiliation[Current address: ]{Oak Ridge National Laboratory,
P.O. Box 2008, Bldg. 5600, MS-6016 Oak Ridge, TN 37831 USA}
\author{Robert W. Boyd}
\affiliation{The Institute of
Optics, University of Rochester, Rochester, New York 14627, USA}


\begin{abstract}
We present a new technique for the detection of two-mode squeezed
states of light that allows for a simple characterization of these
quantum states. The usual detection scheme, based on heterodyne
measurements, requires the use of a local oscillator with a
frequency equal to the mean of the frequencies of the two modes of
the squeezed field.  As a result, unless the two modes are close in
frequency, a high-frequency shot-noise-limited detection system is
needed. We propose the use of a bichromatic field as the local
oscillator in the heterodyne measurements. By the proper selection
of the frequencies of the bichromatic field, it is possible to
arbitrarily select the frequency around which the squeezing
information is located, thus making it possible to use a
low-bandwidth detection system and to move away from any excess
noise present in the system.
\end{abstract}

\pacs{42.50.Dv,42.50.Lc,42.65.-k}

\maketitle

\section{Introduction}

The initial interest in squeezed states was stimulated by the
possibility of increasing the sensitivity of interferometers for
applications such as gravitational wave detection and verification
of relativistic effects. Since then, the field has expanded to other
areas such as quantum optics and atomic physics with the development
of squeezed states of the electromagnetic field
\cite{Slusher85,Shelby86a} and spin squeezed states
\cite{Kitagawa93,Kuzmich99}.

In recent years, multi-mode squeezed states \cite{Ma90,Lo93} have
gained much attention due to the fact that they contain quantum
correlations between the different modes that make up the field. A
specific case of such states is the two-mode squeezed state (TMSS)
whose importance resides in the fact that it is the main source of
continuous-variable entanglement and Einstein-Podolsky-Rosen (EPR)
type correlations. As a result, this quantum state of the
electromagnetic field has found its way into applications such as
continuous-variable teleportation \cite{Braunstein98,Furusawa98},
quantum key distribution \cite{Pereira00,Reid00,Silberhorn02a},
verification of EPR correlations \cite{Reid89,Ou92}, etc.

Although the TMSS has become a fundamental tool for the study of
continuous-variable entanglement, its experimental characterization
still remains a problem. A number of papers
\cite{Kim94,Levandovsky99,Bennink02a} have focused on improving
either the temporal or spatial character of the local oscillator
(LO) used in heterodyne detection in order to optimize the degree of
squeezing measured. However, no attention has been given to the
necessary requirements of the detection system needed for such
measurements.

The usual detection scheme used for the characterization of a TMSS
is based on balanced heterodyne measurements. It requires the use of
a LO with a frequency equal to the mean of the frequencies of the
two modes of the squeezed field. As a result, the squeezing
information is located around the beat note frequency between the LO
and either of the field modes that constitute the squeezed state. In
general the beat note frequency can be arbitrarily large, thus
requiring a high-bandwidth shot-noise-limited detector to perform
the measurement. The combined requirements of high bandwidth and low
noise place difficult constraints on the detection system, since the
electronic noise of the system usually increases as the bandwidth of
the detection system increases.

In this paper we present a simple scheme based on a bichromatic
local oscillator (BLO) that makes it possible to greatly reduce the
bandwidth requirements of the detection system. Characterization of
a TMSS is then much more accessible, for example by using the simple
low-frequency design of Gray \textit{et al.}~\cite{Gray98}.  As will
be shown, by the proper selection of frequencies of the bichromatic
field it is possible to arbitrarily select the frequency around
which the squeezing information is located, thus making it possible
to use a low-bandwidth detection system to characterize a TMSS
source. Since the measurement frequency can be arbitrarily selected,
it is also possible to move away from any excess noise present in
the system.

The rest of the paper is organized as follows. In
Sec.~\ref{BasicSec} we give a general overview of the basic theory
of heterodyne detection for the characterization of a TMSS. Then, in
Sec.~\ref{BLOSec} we introduce the concept of the BLO and show the
advantages and limitations of such a detection technique.

\section{Balanced Heterodyne Detection}\label{BasicSec}

\begin{figure}[h]
\includegraphics{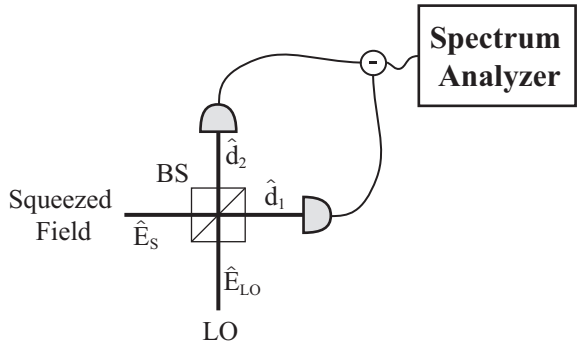}
\caption{\label{balanced} Balanced heterodyne detection scheme used
for the characterization of squeezed light. \textit{Notation}: LO =
local oscillator; BS = beam splitter.}
\end{figure}

The most commonly used technique for the detection of a TMSS is
balanced heterodyne detection.  In general, this technique consists
of combining the squeezed field being measured with a strong LO and
detecting each of the resulting fields with a photodetector, as
shown in Fig.~\ref{balanced}. Combining the two detector signals we
obtain the difference signal and analyze the noise in this signal
with a spectrum analyzer.

The fields after the beam splitter are given by
\begin{eqnarray}
    \hat{d}_{1} & = & t\hat{E}_{S}+r\hat{E}_{LO} \nonumber \\
    \hat{d}_{2} & = & r\hat{E}_{S}+t\hat{E}_{LO},
\end{eqnarray}
where $\hat{E}_{S}$ and $\hat{E}_{LO}$ are the positive frequency
parts of the squeezed and local oscillator fields and $t$ and $r$
are the transmissivity and reflectivity of the beam splitter.  In
general, $t$ and $r$ satisfy the relations $|t|^{2}+|r|^2=1$ and
$t^{*}r=i|rt|$.  In order to have a balanced detection scheme, the
beam splitter must satisfy the condition $|t|=|r|=1/\sqrt{2}$, such
that the difference signal from the balanced heterodyne detection
takes the form
\begin{equation}
    \label{diffsig}
    \hat{I}_{12}=\hatn{d}{1}-\hatn{d}{2}
    =i(\hatnc{E}{S}{LO}-\hatnc{E}{LO}{S}).
\end{equation}
As can be seen from this equation, only the interference terms are
left for the balanced case.

For a TMSS, the field takes the form \footnote{Throughout the paper,
the fields are expressed in units of
$E_{0}=\sqrt{\hbar\omega/2\epsilon_{0}V}$. The frequency difference
between the modes is assumed to be much smaller than the optical
frequency so that $E_{0}$ can be taken as a constant.}
\begin{equation}
    \label{TMSSfield}
    \hat{E}_{S}=\hat{a}_{+}e^{-i\omega_{+}t}+\hat{a}_{-}e^{-i\omega_{-}t}
\end{equation}
and the quadratures are defined according to
\begin{eqnarray}
    \hat{X}&=&\frac{1}{2\sqrt{2}}(\hat{a}_{+}+\hat{a}_{+}^{\dag}+\hat{a}_{-}+\hat{a}_{-}^{\dag})
    \nonumber\\
    \hat{Y}&=&\frac{1}{i2\sqrt{2}}(\hat{a}_{+}-\hat{a}_{+}^{\dag}+\hat{a}_{-}-\hat{a}_{-}^{\dag}).
\label{TMSSquad}
\end{eqnarray}
With these definitions and the properties of the TMSS, the variance
of the quadratures can be shown to be given by \cite{Loudon87}
\begin{eqnarray}
    \var{X}{}& = & \frac{1}{4}\left(e^{-2s}\cos^2\frac{\theta}{2}+e^{2s}\sin^2\frac{\theta}{2}\right)\nonumber\\
    \var{Y}{}& = &
    \frac{1}{4}\left(e^{-2s}\sin^2\frac{\theta}{2}+e^{2s}\cos^2\frac{\theta}{2}\right),
\label{TMSSquadvargen}
\end{eqnarray}
where $s$ is the degree of squeezing and $\theta$ is the squeezing
angle.  As a result, the variance along the major and minor axes of
the squeezing ellipse is given by
\begin{eqnarray}
    \var{X}{}_{min}& = & \frac{1}{4}e^{-2s}\nonumber\\
    \var{Y}{}_{max}& = & \frac{1}{4}e^{2s}.
\label{TMSSquadvar}
\end{eqnarray}

\begin{figure}[hbt]
\includegraphics{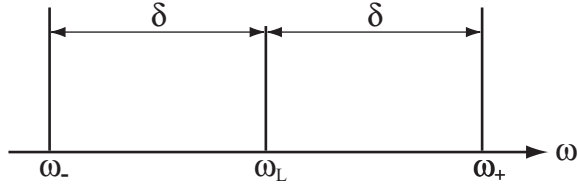}
\caption{\label{freqLO} Frequency components involved in a
heterodyne measurement of a two-mode squeezed state.  The
frequencies of the two modes of the squeezed state are given by
$\omega_{-}$ and $\omega_{+}$ while the frequency of the LO is
represented by $\omega_{L}$. In order to get a measurement that is
independent of time, the frequency of the LO has to be selected
between the frequencies of the two modes of the squeezed state.}
\end{figure}

 In the standard heterodyne technique, the LO is taken to be of
the form
\begin{equation}
    \label{LOfield}
    \hat{E}_{LO}=\hat{b}e^{-i\omega_{L}t}
\end{equation}
and is assumed to be in a coherent state, such that
$\meanI{\hat{b}}=|\beta|e^{i\chi}$. In order to obtain a measurement
that is time independent, the frequency of the LO has to be selected
between the frequencies of the two modes of the squeezed state, as
shown in Fig.~\ref{freqLO}, that is
$\omega_{L}=(\omega_{+}+\omega_{-})/2$.

In the ideal case, the variance of the difference signal,
Eq.~(\ref{diffsig}), is proportional to the noise in the quadratures
of the measured field, thus giving a direct measure of the noise
properties of the squeezed field. That is, in the limit that the LO
is much stronger than the squeezed state, the variance of the
measured signal has the form \cite{Loudon87}
\begin{equation}
\label{heterodynenoise}
    \var{I}{12}=2|\beta|^2\left[e^{2s}\cos^{2}
    \left(\chi-\frac{\theta}{2}\right)
    +e^{-2s}\sin^{2}\left(\chi-\frac{\theta}{2}\right)\right].
\end{equation}
As can be seen from Eqs.~(\ref{TMSSquadvargen}) and
(\ref{heterodynenoise}), the measured quadrature variance can be
selected by changing the phase of the LO, $\chi$. Apart from giving
a signal that is proportional to the noise of the squeezed field,
the balanced heterodyne detection has the additional advantages of
amplifying the measured signal by the strength of the LO, as can be
seen in Eq.~(\ref{heterodynenoise}), and of eliminating both the
quantum and excess noise contributions of the LO.

Once the heterodyne measurement is performed, the squeezing
information is centered around the beat note frequency between the
LO and squeezed field. As mentioned above, for the case of a TMSS,
the frequency of the LO has to be selected such that
$\omega_{L}=(\omega_{+}+\omega_{-})/2$. As a result, the squeezing
information will be centered around the beat note frequency
$\delta=(\omega_{+}-\omega_{-})/2$.  This technique is specially
useful when the frequencies of the modes of the squeezed state are
close together.  However, in general $\delta$ can be arbitrarily
large, requiring as a result a large bandwidth for the
shot-noise-limited detection system.


\section{Bichromatic Local Oscillator}\label{BLOSec}

As seen in Sect.~\ref{BasicSec}, one of the main disadvantages of
the standard balanced heterodyne technique is that the squeezing
information is located around the beat note frequency $\delta$.
Unless the two modes are quite close together, this frequency will
be large making it difficult, and in some cases impossible, to
characterize the TMSS. In order to alleviate the large bandwidth
requirements for the detection system, we propose the use of a
bichromatic field as the LO in the balanced heterodyne technique
described above.  As we will see, by the proper selection of
frequencies of the bichromatic field it is possible to perform
exactly the same measurement as with the standard technique while
making it possible to use a low-bandwidth detection system for
characterizing a TMSS.

\begin{figure}[hbt]
    \includegraphics{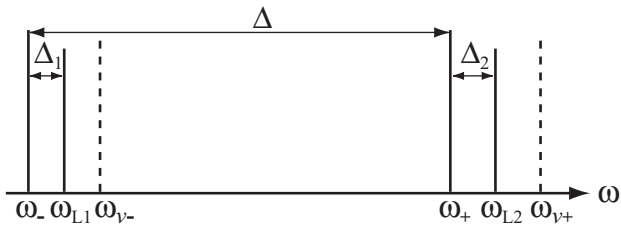}
    \caption{\label{freqBLO} Frequency components for the
    characterization of a two-mode squeezed state using a
    bichromatic local oscillator. The frequencies of the modes of the
    local oscillator are chosen close to each of the modes of the squeezed field.
    It is necessary to include in the analysis the influence of the image bands
    associated with each mode of the squeezed state.  The image bands
    are represented by the dashed lines and are assumed to be in a
    vacuum state.}
\end{figure}

In this new scheme, the local oscillator is taken to be a
bichromatic field of the form
\begin{equation}
    \label{BLOfield}
    \hat{E}_{LO}=\hat{b}_{1}e^{-i\omega_{L1}t}+\hat{b}_{2}e^{-i\omega_{L2}t},
\end{equation}
such that the frequency of each of the local oscillators is taken
close to one of the modes of the squeezed field, as shown in
Fig.~\ref{freqBLO}. That is, $\Delta_{1}=\omega_{L1}-\omega_{-}$ and
$\Delta_{2}=\omega_{L2}-\omega_{+}\ll\Delta=\omega_{+}-\omega_{-}$.
In most cases, such a BLO can easily be generated from the laser
fields used for the generation of the TMSS with the help of either
acousto-optic or electro-optic modulators. Both fields of the BLO
are taken to be in coherent states, such that
$\meanI{\hat{b}_{1}}=\beta_{1}$ and $\meanI{\hat{b}_{2}}=\beta_{2}$.
Using this form for the LO, the variance of the measured signal is
now given by
\begin{eqnarray}
    \label{varBLOTMSS}
    \var{I}{12}&=&-\{\var{E}{S}(\beta_{1}^{2}e^{-i2\omega_{L1}t}+\beta_{2}^{2}e^{-i2\omega_{L2}t}
    +2\beta_{1}\beta_{2}e^{-i(\omega_{L1}+\omega_{L2})t})\nonumber\\
    &&-(\meanI{\hatn{E}{S}}-\meanI{\hat{E}_{S}}\meanI{\hat{E}_{S}^{\dag}})(|\beta_{1}|^{2}+|\beta_{1}|^{2}
    +\beta_{1}\beta_{2}^{*}e^{-i(\omega_{L1}-\omega_{L2})t}+\beta_{1}^{*}\beta_{2}e^{i(\omega_{L1}-\omega_{L2})t})\nonumber\\
    &&+h.c.\}+2\meanI{\hatn{E}{S}},
\end{eqnarray}
where the last term on the right hand side results from the
quantization of the BLO and, as will be seen, can be neglected if
the BLO is taken to be sufficiently strong.

For the general case of the BLO shown in Fig.~\ref{freqBLO} it is
necessary to include the image band for each of the modes of the
squeezed state \cite{Yuen80,Collett87}.  Due to the fact that it is
not possible to distinguish between the positive and negative
frequency beat note signals when looking at the heterodyne signal,
it is necessary to take into account frequencies that lie
symmetrically on either side of the LO, as shown in
Fig.~\ref{freqBLO}.  The mode opposite to the squeezed field is
known as the image band.  As a result, the field that is measured is
not the TMSS given by Eq.~(\ref{TMSSfield}); instead it takes the
form
\begin{equation}
    \label{imageTMSSfield}
    \hat{E}_{S}=\hat{a}_{+}e^{-i\omega_{+}t}+\hat{a}_{v+}e^{-i\omega_{v+}t}+\hat{a}_{-}e^{-i\omega_{-}t}+\hat{a}_{v-}e^{-i\omega_{v-}t},
\end{equation}
where modes $\hat{a}_{v+}$ and $\hat{a}_{v-}$ are the image bands
and are taken to be in vacuum states.  The frequency of these modes
is such that $\omega_{L1}-\omega_{-}=\omega_{v-}-\omega_{L1}$ and
$\omega_{L2}-\omega_{+}=\omega_{v+}-\omega_{L2}$.

From the definition of the image band, it follows that there are two
specific cases in which Eq.~(\ref{imageTMSSfield}) needs to be
modified.  Referring to Fig.~\ref{freqBLO}, the first is when
$\Delta_{1}=\Delta_{2}=0$. In this case there are no image bands,
and the measured field again takes the form given by
Eq.~(\ref{TMSSfield}). This is equivalent to the difference between
using homodyne or heterodyne detection to measure single mode
squeezing \cite{Collett87}. The other case is when
$\Delta_{1}=-\Delta_{2}=\Delta/4$. In this case, the image band of
each LO coincide, so that it is only necessary to take into account
a single vacuum field mode in Eq.~(\ref{imageTMSSfield}).

Using the properties of the TMSS \cite{Loudon87}, the different
parts of Eq.~(\ref{varBLOTMSS}) can be shown to be of the form
\begin{eqnarray}
    \label{TMSSprop}
    \var{E^{\dag}}{S}&=&-2e^{-i\theta}e^{i(\omega_{-}+\omega_{+})t}\sinh s \cosh s \\
    \var{E}{S}&=&-2e^{i\theta}e^{-i(\omega_{-}+\omega_{+})t}\sinh s \cosh s \\
    \meanI{\hatn{E}{S}}+\meanI{\hat{E}_{S}\hat{E}_{S}^{\dag}}-2\meanI{\hat{E}_{S}}\meanI{\hat{E}_{S}^{\dag}}
    &=&4\sinh^{2}
    s+2+\meanI{\hat{a}_{v+}\hat{a}_{v+}^{\dag}}+\meanI{\hat{a}_{v-}\hat{a}_{v-}^{\dag}}.
\end{eqnarray}
Since the image band modes are in the vacuum state,
$\meanI{\hat{a}_{v+}\hat{a}_{v+}^{\dag}}=\meanI{\hat{a}_{v-}\hat{a}_{v-}^{\dag}}=1$;
however, the value of these expressions is not substituted until the
final result in order to see their effect and make the necessary
changes depending on the cases described above.

The BLO technique becomes useful when  $\Delta \gg
\Delta_{1},\Delta_{2}$.  In this case the bandwidth of the detection
system can be designed such that the terms with frequency of the
order of $\Delta$ can be neglected. Under this approximation,
Eq.~({\ref{varBLOTMSS}) and Eq.~({\ref{TMSSprop}) can be combined to
obtain the variance of the difference signal, which can be shown to
be given by
\begin{eqnarray}
    \label{BLOvardiffamp}
    \var{I}{12}&=&(|\beta_{1}|^{2}+|\beta_{2}|^{2})(4\sinh^{2}s+2+\meanI{\hat{a}_{v+}\hat{a}_{v+}^{\dag}}+\meanI{\hat{a}_{v-}\hat{a}_{v-}^{\dag}})\nonumber\\
                &&+4\beta_{1}\beta_{2}e^{-i(\Delta_{1}+\Delta_{2})t}e^{-i\theta}\sinh{s}\cosh{s}
                                          +4\beta_{1}^{*}\beta_{2}^{*}e^{i(\Delta_{1}+\Delta_{2})t}e^{i\theta}\sinh{s}\cosh{s}.
\end{eqnarray}
In order to make the measurement time independent, it is necessary
to select the frequency of the fields of the BLO such that
$\Delta_{1}=-\Delta_{2}$. By making
$\beta_{1}=|\beta_{1}|e^{i\chi_{1}}$ and
$\beta_{2}=|\beta_{2}|e^{i\chi_{2}}$ and assuming that
$|\beta_{1}|=|\beta_{2}|\equiv|\beta|$, we can simplify
Eq.~(\ref{BLOvardiffamp}) to the form
\begin{eqnarray}
\label{BLOnoise}
    \var{I}{12}&=&4|\beta|^2\left[e^{2s}\cos^{2}
    \left(\frac{\chi_1+\chi_2-\theta}{2}\right)
    +e^{-2s}\sin^{2}\left(\frac{\chi_1+\chi_2-\theta}{2}\right) \right. \nonumber\\
    &&\left. +\frac{\meanI{\hat{a}_{v+}\hat{a}_{v+}^{\dag}}+\meanI{\hat{a}_{v-}\hat{a}_{v-}^{\dag}}}{2}
    +\frac{\meanI{\hatn{a}{+}}+\meanI{\hatn{a}{-}}}{2|\beta|^2}\right].
\end{eqnarray}
As described above, the last term in Eq.~(\ref{BLOnoise}) is due to
the quantization of the BLO.  This additional term is a phase
independent noise term that can limit the minimum amount of
squeezing that can be measured. However, it can be neglected when
$|\beta|^{2}\gg(\meanI{\hatn{a}{+}}+\meanI{\hatn{a}{-}})/2$, that
is, the intensity of the BLO is much greater than the intensity of
the TMSS being measured.  This is usually the case for balanced
heterodyne detection.

As mentioned above, due to the image bands, there are three
different cases to consider depending on the values of $\Delta_{1}$
and $\Delta_{2}$. Once the appropriate image bands are taken into
account, the variance in the difference signal is given by
\begin{equation}
\label{varcases}
    \var{I}{12} = \left\{ \begin{array}{ll}
            4|\beta|^2\left[e^{2s}\cos^{2} \left(\frac{\chi_1+\chi_2-\theta}{2}\right)
                +e^{-2s}\sin^{2}\left(\frac{\chi_1+\chi_2-\theta}{2}\right) \right] &\mbox{ if $\Delta_{1}=\Delta_{2}=0$}, \\
            4|\beta|^2\left[e^{2s}\cos^{2} \left(\frac{\chi_1+\chi_2-\theta}{2}\right)
                +e^{-2s}\sin^{2}\left(\frac{\chi_1+\chi_2-\theta}{2}\right) +\frac{1}{2} \right] &\mbox{ if $\Delta_{1}=-\Delta_{2}=\Delta/4$},\\
            4|\beta|^2\left[e^{2s}\cos^{2} \left(\frac{\chi_1+\chi_2-\theta}{2}\right)
                +e^{-2s}\sin^{2}\left(\frac{\chi_1+\chi_2-\theta}{2}\right) +1 \right] &\mbox{
                otherwise}.
       \end{array} \right.
\end{equation}
Except for a scaling factor, the first of these results is exactly
the \ same as the result obtain with the usual balanced heterodyne
technique, Eq.~(\ref{heterodynenoise}). However, the squeezing
information is now centered around DC, making it possible to use a
low-bandwidth detection system.

The other two cases of Eq.~(\ref{varcases}) contain an extra noise
term due to the image bands. This is exactly the situation that
results when using heterodyne detection  for measuring  a single
mode squeezed state \cite{Collett87}.  This extra noise term limits
the amount of squeezing that can be measured.  In the limit of
infinite squeezing, $s\rightarrow\infty$, the second cases in
Eq.~(\ref{varcases}) will give a signal that is 6~dB below the
classical level, while the third case will only give 3~dB below.
However, for the third case it is possible to arbitrarily select the
measurement frequency, thus making it possible to move away from the
$1/f$ noise and any technical noise present in the detection system.

Up to now we have considered only the case in which both fields of
the BLO have exactly the same amplitude.  In practice this is not
always possible, so it is necessary to consider the situation in
which the amplitudes are not properly matched. In order to consider
this case, the amplitudes of the two fields of the BLO are taken to
be $|\beta_{1}|=|\beta|$ and $|\beta_{2}|=|\beta|+\delta\beta$, such
that Eq.~(\ref{BLOvardiffamp}) now takes the form
\begin{eqnarray}
\label{BLOnoiseunbalanced}
    \var{I}{12}&=&4|\beta|^{2}\left\{\left(1+\frac{\delta\beta}{|\beta|}\right)\left[e^{2s}\cos^{2}
    \left(\frac{\chi_1+\chi_2-\theta}{2}\right)
    +e^{-2s}\sin^{2}\left(\frac{\chi_1+\chi_2-\theta}{2}\right) \right. \right. \nonumber \\
    &&\left. \left. +\frac{\meanI{\hat{a}_{v+}\hat{a}_{v+}^{\dag}}+\meanI{\hat{a}_{v-}\hat{a}_{v-}^{\dag}}}{2}\right]
    +\frac{1}{2}\left(\frac{\delta\beta}{|\beta|}\right)^{2}\left(\cosh
    2s+\frac{\meanI{\hat{a}_{v+}\hat{a}_{v+}^{\dag}}+\meanI{\hat{a}_{v-}\hat{a}_{v-}^{\dag}}}{2}\right)\right\}.\nonumber\\
\end{eqnarray}
As can be seen from Eq.~(\ref{BLOnoiseunbalanced}), the imbalance in
amplitudes leads to an extra source of noise. However, this extra
noise term is phase independent and of second order in
$\delta\beta/|\beta|$, so that its contribution can easily be made
negligible.  Thus, to first order, the imbalance in amplitudes has
no effect on the measurement other than an overall scaling factor.

The main advantage gained by using a BLO is that, independently of
the frequency separation between the modes of the squeezed field, it
is possible to characterize a TMSS without the need for a
high-frequency shot-noise limited detector.  Another property of
using a BLO, as can be seen from Eq.~(\ref{BLOnoise}), is that it is
possible to select the measured quadrature variance by changing the
phase of either one of the modes of the BLO, which might have some
application in the detection of correlations in the different
quadratures of the field.

\section{Conclusion}\label{Concl}

We have shown that by using a bichromatic local oscillator in a
heterodyne detection scheme it is possible to use a low-frequency
detection system for the characterization of a TMSS, independently
of the frequency separation between the two modes of the squeezed
state. The BLO required for this type of measurement can easily be
generated from the laser fields used for the generation of the TMSS
with the help of either acousto-optic or electro-optic modulators.
This allows for the use of a simple detection system for the
characterization of any TMSS source. In order to get the same
measurement result as with the standard balanced heterodyne
detection technique, it is necessary to select the frequencies of
the BLO to coincide with the frequencies of the two modes of the
squeezed state. However, analogous to the case of single mode
squeezing, it is possible to arbitrarily select the detection
frequency at the expense of extra noise.  This freedom to select the
desired detection frequency makes it possible to move away from the
$1/f$ noise or any technical noise present in the detection system.
In principle, it is possible to extend this idea to the general case
of multi-mode squeezed states.

\begin{acknowledgments}
We would like to acknowledge useful discussions with Claude Fabre
and Frank Narducci. This work was supported in part by the National
Science Foundation.
\end{acknowledgments}


\end{document}